\documentclass[%
 reprint,
 amsmath,amssymb,
 aps,
]{revtex4-2}

\usepackage[pdfborder={0 0 0}]{hyperref}
\usepackage{graphicx}
\usepackage{dcolumn}
\usepackage{bm}
\usepackage{xcolor}
\usepackage{soul}
\usepackage[utf8]{inputenc}
\usepackage{hyperref, graphicx, bbold, subcaption, ragged2e}

\usepackage{color}

\graphicspath{{./}{./figures/}}

\def\vE{ {\boldsymbol{E}} }

\def\vm{ {\boldsymbol{m}} }

\def\vp{ {\boldsymbol{p}} }

\def\vx{ {\boldsymbol{x}} }

\newcommand{\bxi}{\boldsymbol{\xi}}

\newcommand{\bsg}{\boldsymbol{\sigma}}

\newcommand{\beq}{\begin{equation}}
\newcommand{\eeq}{\end{equation}}

\newcommand{\uvec}{\mathbf{e}}

\newcommand{\up}{\uparrow}

\def\bra#1{\left\langle#1\right|}
\def\ket#1{\left|#1\right\rangle}
\def\braket#1#2{\left\langle#1\middle|#2\right\rangle}

\begin{document}

\title{Spin preservation in screw-symmetric molecules}

\author{Jonas Bloch}
\email{jonas.bloch@u-pariscite.fr}
\author{Fedor Baranov}
\author{Maxim Breitkreiz}
\email{breitkr@physik.fu-berlin.de}
\affiliation{Dahlem Center for Complex Quantum Systems, Halle-Berlin-Regensburg Cluster of Excellence CCE, and Fachbereich Physik, Freie Universit\"at Berlin, Arnimallee 14, 14195 Berlin}

\date{\today}

\begin{abstract}
In electronic transport through long molecules, spin is expected to be preserved only when the dwell time is much shorter than the characteristic spin-mixing timescale $\hbar/\Delta$, where $\Delta$ is the magnitude of a spin-dependent potential, such as spin-orbit coupling. We show that, in molecules featuring discrete screw symmetry, spin preservation can be enhanced far beyond this timescale owing to strong spin separation in quasi-momentum. This spin fidelity in long molecules is consistent with chirality-induced spin selectivity (CISS), suggesting spin-dependent transport in long, chiral molecules with amplified spin-splitting mechanisms. We provide analytical derivation of the enhanced spin preservation and test it on tight-binding models, which confirm that the effect gradually weakens when the screw symmetry is broken or changes from discrete to continuous. Furthermore, we perform transport simulations to show that a strong magnetoresistance trace of symmetry-protected spin fidelity emerges in a spin-valve setup with two magnetic leads, which we propose as an experimentally accessible signature.
\end{abstract}

\maketitle

\section{Introduction}

Since the pioneering works of Ray et al. \cite{ray_asymmetric_1999}, electron-spin dynamics in chiral molecules has attracted considerable attention. Over the past decades, it was consistently reported that chirality induces a remarkable spin dependence upon electron transport through the molecule, with the spin quantized parallel to the electron's velocity --- a phenomenon now known as \emph{Chirality Induced Spin Selectivity} (CISS) \cite{PhysRevB.68.115418, PhysRevLett.96.036101, wei_molecular_2006, gohler_spin_2011, kiran_helicenesnew_2016, mishra_length-dependent_2020, bloom_chiral_2024, rohmer_chiral_2025}. Furthermore, theoretical work has sought to model this interplay between chirality and spin, and much has yet to be understood \cite{guo_spin-selective_2012, diaz_spin_2018, evers_theory_2022, chiesa_many-body_2024}. Spin-orbit coupling (SOC) is believed to be a necessary ingredient, despite being weak in organic molecules that make up the majority of CISS devices. Indeed, SOC can be used to successfully predict a spin filtering in helical molecules, albeit orders of magnitude weaker than the one observed. This implies that, if SOC is indeed at the origin of CISS, an amplification mechanism must exist to attain the measured levels, for example through coupling to phonons \cite{fransson_vibrational_2020} or through close-to-barrier tunneling \cite{Baranov2026}.

Most of these efforts focus on the transmission difference between opposite input spins, regardless of whether these spins change during the traversal through the molecule: in most experiments, only one terminal is spin polarized. Moreover, experiments probing the evolution of spin-polarization over electron transport are notoriously difficult to carry out on organic molecules \cite{sanvito_molecular_2011}. This is due to their light atoms causing weak spin-orbit interaction, which also happens to be a key mechanism underlying many spintronic detection methods. As an example, the spin-splitting induced by SOC goes as high as 340 meV in GaAs, whereas it barely reaches 13 meV in diamond \cite{sanvito_molecular-spintronics_2006}, which is itself a coarse upper bound to what can be found in helical organic molecules \cite{evers_theory_2022}. Therefore, little is known about spin-flipping processes inside the molecule, although some results have already been established. In a previous work, macroscopic rate equations fed with experimental data were used to reconstruct the evolution of the spin populations over the traversal, which suggested that spin flips practically never happen \cite{nurenberg_evaluation_2019}. Further results showed that different input spins yield separate conduction voltage thresholds, confirming that spin flips are marginal \cite{mishra_spin-filtering_2020}, and that spin-chirality interactions mainly consist of spin-preserving scattering processes. A deeper understanding of spin preservation and proposals for experimental setups capable of probing spin dynamics throughout molecular transport is therefore crucial for advancing our understanding of CISS and the broader interplay between chirality and spin dynamics.

In this work, we show that screw symmetry provides a robust mechanism for preserving an electron’s spin in transport across a broad class of helical molecules. We study the dynamics of a single electron in the presence of a generic spin-splitting potential that includes SOC. We demonstrate that screw symmetry strongly suppresses spin-flipping processes and naturally explains the near absence of spin flips observed experimentally. Furthermore, we show that this could give rise to remarkable magnetoresistive properties in spin-valve setups, where both leads are magnetic and arranged in an antiparallel magnetization alignment.

Our analysis assumes that the spin-splitting energy remains realistically small compared to the other energy scales. Previously, the weakness of spin-flip scattering has been attributed to this smallness \cite{li_too_2025}, which, however, sits uneasily alongside the widely discussed need for mechanisms that amplify the effects of spin splitting to account for CISS.
Instead, we connect the suppression of spin flips with the presence of screw symmetry. {This symmetry, which belongs to the family of nonsymmorphic symmetries, is known to create persistent spin textures  \cite{tao_persistent_2018, ji_symmetry-protected_2022, kilic_universal_2025, shang_light-induced_2024} which are considered as a route to achieve long spin lifetimes \cite{bernevig_exact_2006, PhysRevB.90.201306}. Here we derive extended spin lifetimes explicitly for the case of screw-symmetric molecules with weak spin splitting.
Furthermore, we show that in the presence of screw symmetry, an effective spin splitting arises that is perfectly aligned with the screw axis, even without being so locally. The resulting spin-degeneracy lifting is a necessary ingredient for spin-polarized transport effects  \cite{fransson_chiral_2025, Baranov2026} and the formation of triplet pairs \cite{FranssonNaaman25}.}


The remainder of this paper is organized as follows: in Sec.~\ref{sec:screw-symmetry-model}, a model of a screw-symmetric molecule is introduced alongside a model for the dynamics of a single electron moving through it. In Sec.~\ref{sec:spin-flip-derivation}, a derivation of the quenched spin flips is given, to be tested in numerical toy models in Sec.~\ref{sec:numerical-models}. In Sec.~\ref{sec:experimental-probes}, possible experimental probes of the effect are discussed. A discussion on the limitations of the assumptions underlying the proposed effect is given in Sec.~\ref{sec:discussion}.

\section{Screw-symmetric molecule}
\label{sec:screw-symmetry-model}

We consider a molecule described by a periodic potential $V({\vx})$, infinite or terminated with periodic boundary conditions in the direction of $\uvec_3$, in the canonical basis $(\uvec_1, \uvec_2, \uvec_3)$ for the 3D space. We also consider the existence of a spin-dependent potential $\hat H_{\text{sp}}$, sufficiently weak to be treated in first order perturbation theory.

\begin{figure}
  \begin{center}
    \includegraphics[height=0.3\textwidth]{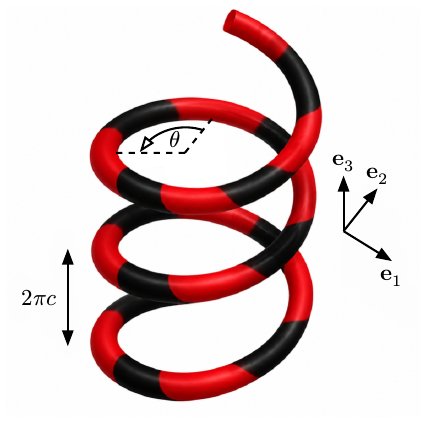}
  \end{center}
  \caption{\justifying Schematic illustration of a screw-symmetric helical object with screw angle $\theta = \pi/2$, chirality $\chi = -1$, and pitch $2\pi c$, encoded in the periodic color texture alternating between black and red.}\label{fig:helix-example}
\end{figure}

To define the screw symmetry of the molecule, we introduce the screw sense (chirality) $\chi \in \{-1, 1\}$, the pitch $2\pi c > 0$, and the screw angle $\theta > 0$, as illustrated in Fig.~\ref{fig:helix-example}. The \emph{screw affine map} of parameters $c, \chi, \theta$ is defined by
\begin{equation}
  \mathfrak s_{c, \chi, \theta} = \vx \mapsto U_{\chi \theta} {\vx} + \theta c \uvec_3,
\end{equation}
where $U_{\chi \theta}$ is the matrix rotating a vector around $\uvec_3$ by an angle $\chi\theta$. We postulate the existence of parameters $c, \chi, \theta$ with $\theta \notin 2 \pi \mathbb Z$ such that
\begin{equation}
\forall {\vx} \in \mathbb R^3, ~ V\left(\mathfrak s_{c, \theta, \chi} ({\vx})\right) = V({\vx}).
\end{equation}
The screw symmetry belongs to nonsymmorphic symmetries, which in general consist in a combination of a fractional translation of the crystal lattice and a point group symmetry \cite{10.1093/oso/9780199582587.001.0001}. For screw symmetry, the latter is a rotation. We emphasize that $\theta/2\pi$ must not be an integer, since otherwise the screw symmetry reduces to a regular translational symmetry and is no longer nonsymmorphic. We assume that this symmetry applies to any position-dependent quantity related to the molecule, such as the electric field $\vE$ associated with the molecular potential, $\vE (\mathfrak s_{c, \chi, \theta} (\vx)) = U_{\chi\theta} \vE (\vx)$.

Molecules with screw symmetry are often helical, but the two notions are not equivalent. A molecule can be helical without screw symmetry, such as a DNA strand with a non-periodic sequence. Conversely, a screw-symmetric molecule is not necessarily helical: if $\theta = \pi$, then $\mathfrak s_{c, 1, \theta} = \mathfrak s_{c, -1, \theta}$, the molecule described can be achiral, and therefore also non-helical.

Upon quantization, $\vx$ becomes an operator $\hat{\vx}$ and we can represent the screw symmetry with a unitary operator,
\begin{align}
\mathfrak s_{c, \chi, \theta}(\hat{\vx})  &= \hat{S}_0^\dagger \hat{\vx} \hat S_0, \ \ 
  \hat{S}_0 \equiv e^{i \theta \left(c {\hat p_3} + \chi {\hat J_3}\right)/\hbar}, 
\end{align}
where $\hat p_3$ and $\hat J_3$ are the linear and angular momenta operators along $\uvec_3$. Adding the spin degree of freedom using the vector of Pauli operators $\hat{\bsg}$, the screw operator generalizes to 
\begin{equation}
\hat S = \hat S_0 e^{i \hat \sigma_3 \chi \theta / 2}.
\end{equation}
The distinction between $\hat{S}$ and $\hat{S}_0$ will turn out to be useful for diagonalizing the full Hamiltonian including the spin-splitting interaction. The dynamics of the electron are initially described by the (spin-independent) Hamiltonian
\begin{equation}
\hat H_0 = \frac{\hat p^2}{2 m_e} + V(\hat{\vx}),
\end{equation}
where $m_e$ is the electron's (effective) mass. Since the operators $\hat H_0, \hat S, \hat \sigma_3$ commute pairwise, we consider the eigenstates simultaneously indexed by the eigenvalues of $\hat S$ and $\hat \sigma_3$. We introduce a wavenumber $k \in (-\pi, \pi]$ such that $\exp(ik)$ is the eigenvalue of $\hat S$, and $\sigma \in \{-1, 1\}$ is the eigenvalue of $\hat \sigma_3$. Thus, we obtain the eigenstates $\ket{\psi_{k, \sigma, n}}$ with $\hat H_0 \ket{\psi_{k, \sigma, n}} = E_{k, \sigma, n} \ket{\psi_{k, \sigma, n}}$. Since the following analysis does not depend on the band index $n$, we drop it.

\begin{figure}[b]
  \begin{center}
    \includegraphics[height=0.23\textwidth]{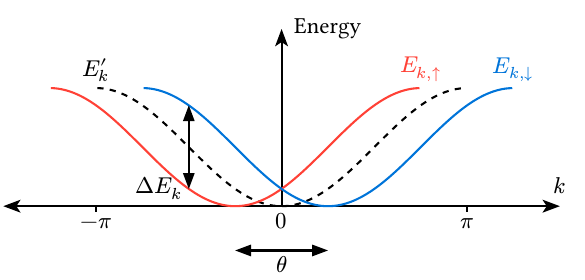}
  \end{center}
  \caption{\justifying Schematic illustration of the spin-dependent band shift by $\chi\theta$ arising from the screw geometry (here, $\chi = -1$ and $\theta = \pi/2$). A strong energy splitting, $\Delta E_k = E_{k, \uparrow} - E_{k, \downarrow} \ne 0$ arises for most values of $k$. The only values of $k$ with a weak splitting correspond to the intersection points of the curves for $E_{k, \uparrow}$ and $E_{k, \downarrow}$.}\label{fig:brillouin-split}
\end{figure}

Note that, despite $\hat H_0$ not involving the Pauli operators, the energy $E_{k, \sigma}$ does depend on $\sigma$. The reason is that $\hat S$ involves $\hat \sigma_3$, and thus so does the Bloch Hamiltonian. Naturally, we could have used $\hat S_0$ instead of $\hat S$, with which we would have obtained a spin-independent dispersion relation $E'_k$. The relation between the two expressions reads
\begin{equation}
E_{k, \sigma} = E'_{k - \sigma \chi \theta / 2}
\end{equation}
and is illustrated in Fig.~\ref{fig:brillouin-split}. 

\section{Spin-flipping scattering}
\label{sec:spin-flip-derivation}

{Crystal symmetries can strongly constrain the spin polarization of single-electron eigenstates. Tao and Tsymbal \cite{tao_persistent_2018} showed that nonsymmorphic symmetry enforces a persistent spin texture around certain high-symmetry points of the Brillouin zone: at momenta invariant under a mirror (or glide mirror), the spin expectation $\langle \hat \bsg \rangle$ is perpendicular to the mirror plane. In two-dimensional materials with an in-plane mirror, this holds over the entire Brillouin zone \cite{ji_symmetry-protected_2022}. Analogously, a screw axis aligns $\langle \hat \bsg \rangle$ with its axis \cite{shang_light-induced_2024, kilic_universal_2025}. Here we consider a discrete screw symmetry about $\uvec_3$ with a weak spin-splitting potential and show that every eigenstate has not only $\langle\hat \bsg\rangle$ aligned with $\uvec_3$, but also that $|\langle \hat \sigma_3 \rangle|\approx1$. Hence $\hat \sigma_3 $ is approximately conserved, such that a spin injected along $\uvec_3$ barely precesses, and precession-driven spin relaxation is strongly suppressed.}


We consider the effect of a spin-dependent potential, which we write as $\hat H_{\text{sp}} = \hat{\vm} \cdot \hat{\bsg}$, where $\hat{\vm}$ is a screw-symmetric spin-splitting field, meaning that it transforms under $\hat S$ like $\hat S^\dagger \hat \vm \hat S = U_{\chi \theta} \hat \vm$.
It can include 
SOC, in which case $\hat{\vm} = \frac{\mu_B}{2 m_e c_0^2} (\hat{\vp} \times {\vE} (\hat{\vx}) - {\vE} (\hat{\vx}) \times \hat{\vp})$ with $\vE$ the electric field associated with the molecule potential, $\mu_B$ the Bohr magneton, and $c_0$ the speed of light in vacuum. Additionally, the spin-splitting field can include a magnetic field induced by currents inside of the molecule \cite{fourcurrent,Bro2025,Baranov2026}. 

Using that $\hat\bsg$ also transforms under $\hat S$ like $\hat S^\dagger \hat{\bsg} \hat S = e^{-i \hat \sigma_3 \chi \theta / 2} \hat{\bsg} e^{i \hat \sigma_3 \chi \theta / 2} = U_{\chi\theta} \hat{\bsg}$, the spin-dependent potential remains invariant under screw symmetry, 
\begin{equation}
\hat S^\dagger \hat H_\text{sp} \hat S = (U_{\chi \theta} \hat{\vm}) \cdot (U_{\chi \theta} \hat{\bsg}) = \hat{\vm} \cdot \hat{\bsg} = \hat H_\text{sp}.
\end{equation}
Note that while the full Hamiltonian $\hat H = \hat H_0+\hat H_\text{sp}$ is thus still screw-symmetric, the spin is no longer preserved if $\hat \vm$ is not parallel to $\uvec_3$ since $[\hat{\sigma_3},\hat H_\text{sp}]\neq 0$. 

Now, we will focus on the probability of $\hat H_\text{sp}$ to induce a spin-flip, that is, a transition $(k, \sigma) \to (k', -\sigma)$
Since $[\hat H_{\text{sp}},\hat S]=0$, the eigenvalue $e^{ik}$ of $\hat S$ is conserved, hence $k' = k$. Perturbation theory dictates that a transition is likely only when the energy difference of the two states is on the order of the perturbation \cite{cohen-tannoudji_quantum_2020}, i.e., 
\begin{equation}
E_{k, \sigma} \approx E_{k, -\sigma},
\end{equation}
or, expressed in terms of the spin-independent dispersion relation, $E'_{k - \theta/2} \approx E'_{k + \theta/2}$. If $\theta$ is far from being a multiple of $2 \pi$, the band splitting is on the order of the band width for most quasi-momenta. The only quasi-momenta at which this condition is guaranteed to be satisfied are the time-reversal-invariant momenta $k=0$ and $k=\pm\pi$, where the bands cross according to Kramers' degeneracy, as is seen in Fig.~\ref{fig:brillouin-split}. Other unprotected (near) degeneracies may occur accidentally. In any case, the range of energies with (near) spin degeneracy at which spin flips can occur make up only for a marginal part of the total spectrum.

We will now provide more substance to this argument by deriving the eigenstates analytically. Using the $\hat S$-symmetry, we can express a complete Bloch Hamiltonian $\hat H = \hat H_0 + \hat H_\text{sp}$ in the spin-state basis in the form
\begin{equation}
\forall k, ~ h(k) = \begin{pmatrix}
  E_{k, \uparrow} + u_{k, \uparrow} & t_k^* \\
  t_k & E_{k, \downarrow} + u_{k, \downarrow}
\end{pmatrix},
\end{equation}
where $u_{k, \sigma} = \bra{\psi_{k, \sigma}} \hat H_\text{sp} \ket{\psi_{k, \sigma}}$ and $t_k = \bra{\psi_{k, \downarrow}} \hat H_\text{sp} \ket{\psi_{k, \uparrow}}$. In order of magnitude, we expect $|u_{k, \sigma}| \sim |t_k| \ll |E_{k, \sigma}|$. We can diagonalize it and identify its eigenvectors $(c^\pm_{k, \uparrow}, c^\pm_{k, \downarrow})^T$ indexed by $\pm$. We focus our attention on the ratio $c^\pm_{k, \downarrow}/c^\pm_{k, \uparrow}$, which is sufficient to evaluate the spin-polarization of the state, namely $(1 - |c^\pm_{k, \downarrow}/c^\pm_{k, \uparrow}|^2)/(1 + |c^\pm_{k, \downarrow}/c^\pm_{k, \uparrow}|^2)$. We have, explicitly,
\begin{equation}
\label{eq:eigvecs-coefficients-ratio}
\frac{c^\pm_{k, \downarrow}}{c^\pm_{k, \uparrow}} = \frac{-\frac{\Delta E_k + u_{k, \uparrow} - u_{k, \downarrow}} 2 \pm \sqrt{\left(\frac{\Delta E_k + u_{k, \uparrow} - u_{k, \downarrow}} 2\right)^2 + |t_k|^2}}{t_k^*},
\end{equation}
where $\Delta E_k = E_{k, \uparrow} - E_{k, \downarrow}$. If we had a simple translational symmetry with $\theta \in 2 \pi \mathbb Z$, we would have $\Delta E_k = 0$, and thus this ratio would be of the order of unity: the eigenstates would have an arbitrary spin-polarization along $\uvec_3$. Now, with a proper screw symmetry, the ratio is dominated by $\Delta E_k$. We eventually find that, if $\Delta E_k > 0$,
\begin{align}
    \left|\frac{c^+_{k, \downarrow}}{c^+_{k, \uparrow}}\right| &\mathop\sim_{|\Delta E_k| \gg |t_k|, |u_{k, \sigma}|}  \frac{|t_k|}{\Delta E_k} \ll 1\text{ and }
\label{eq:molecular-orbitals-coefs-ratios1}
\\
  \left|\frac{c^-_{k, \downarrow}}{c^-_{k, \uparrow}}\right| &\mathop\sim_{|\Delta E_k| \gg |t_k|, |u_{k, \sigma}|} \frac{\Delta E_k}{|t_k|} \gg 1,
\label{eq:molecular-orbitals-coefs-ratios2}
\end{align}
with the opposite conclusions if $\Delta E_k < 0$. {As a consequence, and since the eigenstates take the form $\ket{\psi_{k, \pm}} = c_{k, \uparrow}^\pm \ket{\psi_{k, \uparrow}} + c_{k, \downarrow}^\pm \ket{\psi_{k, \downarrow}}$, we obtain that $\langle \hat\sigma_3 \rangle = \bra {\psi_{k, \pm}} \hat \sigma_3 \ket{\psi_{k, \pm}}$ approaches $1$ or $-1$, since either $\ket{\psi_{k, \uparrow}}$ or $\ket{\psi_{k, \downarrow}}$ dominates.} In other words, $\sigma$ effectively becomes a good quantum number{, and is therefore well preserved over time, independently on the local direction of the spin-splitting field $\hat \vm$.}

To summarize this part, screw symmetry dictates a conservation of the spin-shifted quasi-momentum $k$, for most of which the spins are split by a large energy difference on the order of the band width. 

An important remark is that this result does not generally hold {if $\hat H_{\text{sp}}$ and $\hat H_0$ are of comparable magnitudes or if the screw symmetry is continuous}, i.e., when the characteristic angle $\theta$ of the screw symmetry is small. In the limit $\theta \to 0^+$, the condition $E'_{k - \theta/2} \approx E'_{k + \theta/2}$ is trivially satisfied at all momenta{, such that $\Delta E_k$ approaches $0$ and the spin is no longer conserved according to Eq.~\eqref{eq:eigvecs-coefficients-ratio}. More precisely, the conservation fails when $\Delta E_k = E'_{k - \chi \theta / 2} - E'_{k + \chi \theta / 2} \approx -\chi \theta (\mathrm d E'_k / \mathrm d k)$ becomes comparable to $u_{k, \sigma}$ or $t_k$, i.e. when $\theta$ is of the order of $t_k/(\mathrm d E'_k / \mathrm d k)$. For generic $\mathrm d E'_k / \mathrm d k$ on the order of the bandwidth, the conservation should thus hold as long as 
\begin{equation}
\theta \|\hat H_0\| \gg \|\hat H_\text{sp}\|,
\label{eq:theta-condition-highscrew}
\end{equation}
where $\|\cdot\|$ represents the norm of an operator, giving an estimate of its magnitude, as numerically demonstrated in Sec.~\ref{subsec:transition-to-continuum} below. When this happens, a spin's lifetime becomes finite and of the order of $\hbar / \|\hat H_\text{sp}\|$ (for generic $\hat\vm$  not aligned with $\uvec_3$).

It is worth noting that these limitations do not affect the conservation of $k$. Thus, in the $\hat H_\text{sp}$-induced transitions $(p, \sigma) \to (p + \theta \chi \sigma, -\sigma)$ (where $p = k - \theta \chi \sigma / 2$ is the momentum associated with the spinless screw operator $\hat S_0$), the spin flip is tight to a simultaneous momentum change. This may lead to other spin-dependent transport properties specific to screw-symmetric molecules, which we will not further explore in this work.}

\section{Toy tight-binding models}
\label{sec:numerical-models}

We now test the spin preservation due to screw symmetry numerically on tight-binding models. To this end, we define the spin \emph{fidelity} to quantify the preservation of spin over time, and probe the transition from high to low fidelity when the discrete screw symmetry is destroyed.

In the first model, we test the significance of the discrete screw symmetry as opposed to a continuous one, as evoked at the end of Section~\ref{sec:spin-flip-derivation}. In the second model, we test the significance of the screw symmetry by breaking it while keeping the translational symmetry and the helical geometry intact.

\subsection{Transition to a continuous screw symmetry}
\label{subsec:transition-to-continuum}

As discussed above, if $\theta$ goes to $0^+$, the enhanced spin preservation could vanish. To test this, we consider a tight-binding model of a helical strand of ions following Ref.\ \cite{fransson_vibrational_2020}. The molecule consists of $N \in \mathbb N^*$ coils, $m \ge 2$ identical ions per coil, and the Hamiltonian $\hat H= \hat H_0+\hat H_\mathrm{sp}$ is given by
\begin{align}
\begin{split}
  \hat H_0 &= -t_0 \sum_{j = 0}^{m N - 1} (\ket{j + 1}\bra{j} + \ket{j}\bra{j+1}) \otimes \operatorname{Id}_\sigma, \\
      {\hat H_\text{sp}} &= i \lambda \sum_{j=0}^{m N - 1} (\ket{j + 2}\bra{j} - \ket{j}\bra{j + 2}) \otimes \bxi_j \cdot \hat{\bsg}.
\end{split}
\label{eq:hamiltonian-helical-strand}
\end{align}
Here, $t_0 > 0$ is the normal hopping amplitude, $\lambda \in \mathbb R$ is the (weaker) hopping associated with SOC, $\ket j$ the spinless Wannier orbitals, $\operatorname{Id}_\sigma$ the identity operator acting on spin space, and $\bxi_j$ hopping directions associated with SOC (with $\|\bxi_j\| = 1$). In Ref.\ \cite{fransson_vibrational_2020}, $\bxi_j$ is computed as the cross product of two consecutive bond vectors. Here, we can generalize this by assuming only $\bxi_{j+1} = U_{2 \pi \chi / m} \bxi_j$, taking $\bxi_0$ as a parameter of the model and computing the other $\bxi_j$'s from it. 

In order to estimate the preservation of the spin of a molecular orbital over time, we consider an electron in a state $\ket{\psi(t = 0)} = \ket{\psi_{k, \sigma}}$ and probe the evolution of its spin as a function of time by solving Schrödinger's equation. More specifically, we compute the spin \emph{fidelity} reached over a time $T$, which we define as
\begin{equation}
\mathcal F = \min_{0 \le t \le T} \sum_{k'} |\braket{\psi (t)}{\psi_{k', \sigma}}|^2.
\end{equation}
Under screw symmetry, the only non zero term in the sum is $\left|\braket{\psi(t)}{\psi_{k, \sigma}}\right|^2$ because $k$ is preserved. This minimum is well-defined, as the trajectory of $\ket{\psi(t)}$ in this system is a Rabi cycle between $\ket{\psi_{k, \uparrow}}$ and $\ket{\psi_{k, \downarrow}}$. The fidelity is thus related to the amplitude of that precession. 

In the model of Eq.\ \eqref{eq:hamiltonian-helical-strand}, the screw angle is set by the number of ions per coil as $\theta = 2 \pi / m$. A continuous screw symmetry is thus realized in the limit $m \to +\infty$. We therefore expect the fidelity to be close to $1$ when the screw symmetry is effective with low $m$ and to go down as $m$ increases.

\begin{figure}
  \begin{center}
    \includegraphics[width=0.48\textwidth]{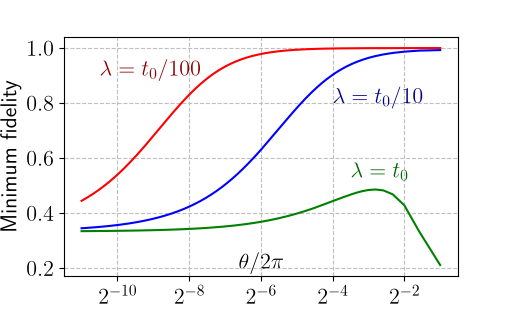}
  \end{center}
  \caption{\justifying Spin fidelity $\mathcal F$ computed at $k = \pi / 3$ and initial spin $\uparrow$. One observes a loss in fidelity when $\theta = 2 \pi / m$ decreases, as the molecule approaches a continuum of ions. Here, $t_0 = 1\,\text{eV}, \bxi_0 = -(1, 1, 1)^T / \sqrt 3, \chi = 1$ and $\lambda$ is indicated on the plots.}\label{fig:transition-to-continuum}
\end{figure}

The calculated fidelity, plotted in Fig.~\ref{fig:transition-to-continuum}, confirms this expectation{, as the loss in fidelity starts when $m$ reaches $2^6$ (resp. $2^3$) when $\lambda = t_0 / 100$ (resp. $t_0/10$), i.e. $\theta t_0 \sim 10 \lambda$. Furthermore, when $\lambda$ is not small compared to $t_0$ (in the case $\lambda = t_0$), spin fidelity is never enhanced}. We refer to Appendix~\ref{apx:continuous-screw-symmetry} for the details of the calculation underlying the plot. Furthermore, Fig.~\ref{fig:transition-to-continuum} shows that the spin flips are also quenched at $m = 2$, which is the special case of a screw-symmetric but non-chiral molecule.
Note that, for the specific choice of $\bxi_0$ corresponding to Ref.~\cite{fransson_vibrational_2020}, all $\bxi_j$'s would converge to $\uvec_3$ as $m \to +\infty$, and thus $\mathcal F$ should remain high as the spin of the eigenstates will be quantized along that direction. Ref.\ \cite{diaz_spin_2018} is another example where the enhanced spin fidelity can survive the continuum limit under additional specific conditions.

\subsection{Broken screw symmetry}
\label{subsec:transition-from-translational}

\begin{figure*}
  \begin{center}
    \includegraphics[width=0.95\textwidth]{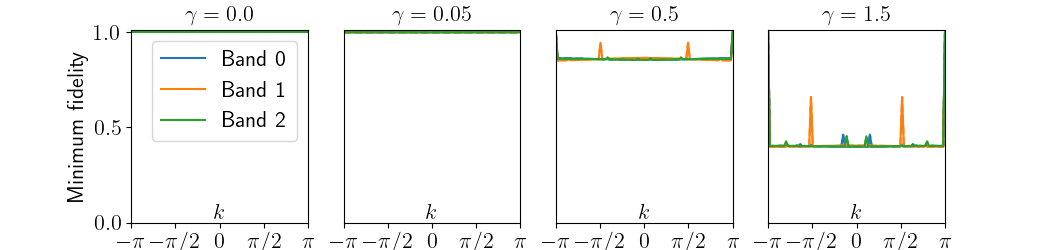}
  \end{center}
  \caption{\justifying Spin fidelity $\mathcal F$ computed against $k$ with initial spin $\uparrow$ for different values of $\gamma$, using $\mathcal F = \min_{0 \le p \le 100} \sum_{n=1}^3 \left|\braket{\psi \left(t = 200 p \frac \hbar {t_0}\right)}{\psi_{k, n, \uparrow}}\right|^2$. The screw angle is fixed to $\theta=2\pi/3$. Other parameters are identical to those in Fig.~\ref{fig:transition-to-continuum}.}\label{fig:transition-from-translational}
\end{figure*}

We continue to consider the tight-binding model of Sec.~\ref{subsec:transition-to-continuum} with the number of sites in one coil fixed to $m = 3$. However, we break the screw symmetry by modifying the hopping amplitudes around each atom in $\hat H_0$, which now reads 
\begin{align}
\begin{split}
  \hat H_0 &= - t_0 \sum_{j=0}^{N - 1} [(1 + \gamma) \ket{3j + 1} \bra{3 j} + (1 - \gamma)  \\
  &     \times\ket{3j + 2}\bra{3j + 1}    + \ket{3j + 3} \bra{3 j + 2}]\otimes \operatorname{Id}_\sigma  + \ \mathrm{h.c.},
\end{split}
\label{eq:hamiltonian-imbalanced-helical-strand}
\end{align}
where $\gamma$ is some real parameter that changes the hopping amplitudes. The corresponding Bloch Hamiltonian is derived in Appendix~\ref{apx:translational}. For $\gamma = 0$, the screw symmetry is restored with $\theta = 2 \pi / 3$. For $\gamma \ne 0$, the screw symmetry is broken and only the usual translational symmetry with a periodicity of three sites is left. 

Numerical results shown in Fig.~\ref{fig:transition-from-translational} confirm that the fidelity increases gradually as the screw symmetry is restored when $\gamma \to 0$. This highlights two important aspects of the result: (i) a generic one-dimensional molecule only equipped with translational symmetry does not necessarily preserve the spin, which is a trait specific to screw symmetry (as can be seen in the case $\gamma = 1.5$), and (ii) a molecule without screw symmetry can yield a good preservation of the electron's spin if the screw symmetry is broken only weakly. In our parametrization, weakly broken screw symmetry is for $|\gamma| \ll 1$, which appears to extend up to relatively large $\gamma \lesssim 0.5$, for which the fidelity is still at least $85 \% $. This latter observation could matter in CISS experiments, given that many helical molecules, such as DNA, are not exactly screw-symmetric.

\section{Possible experimental probes}
\label{sec:experimental-probes}

\subsection{Spin-valve setup}

\begin{figure*}
    \centering
    \begin{subfigure}[t]{0.5\textwidth}
        \centering
        \includegraphics[width=\textwidth]{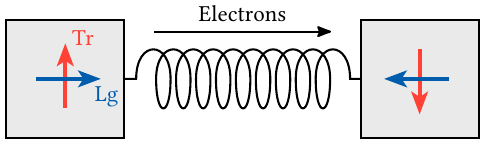}
        \subcaption{}
        \label{subfig:spin-valve-antiparallel-sketch}
    \end{subfigure}
    \\
    \begin{subfigure}[t]{0.49\textwidth}
        \centering
        \includegraphics[width=\textwidth]{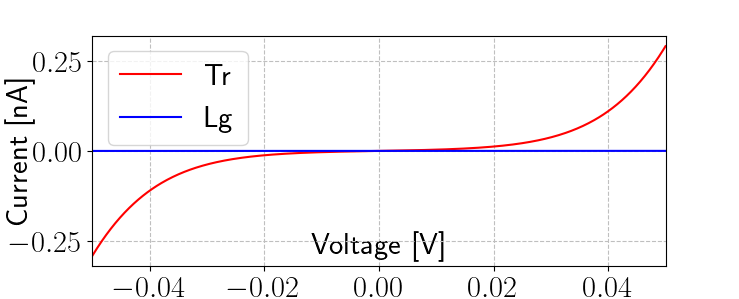}
        \subcaption{}
        \label{subfig:spin-valve-antiparallel-simulation-highscrew-vs-voltage}
    \end{subfigure}
    \begin{subfigure}[t]{0.49\textwidth}
        \centering
        \includegraphics[width=\textwidth]{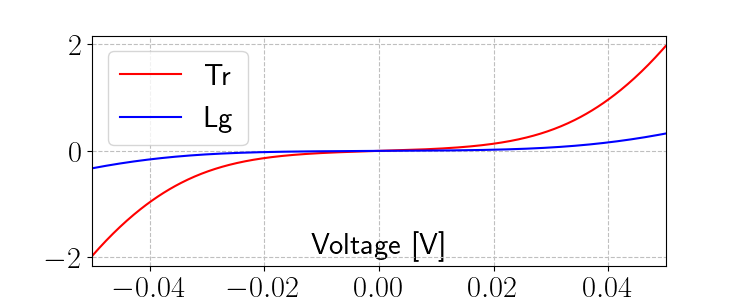}
        \subcaption{}
        \label{subfig:spin-valve-antiparallel-simulation-noscrew-vs-voltage}
    \end{subfigure}
    \\
    \begin{subfigure}[t]{0.49\textwidth}
        \centering
        \includegraphics[width=\textwidth]{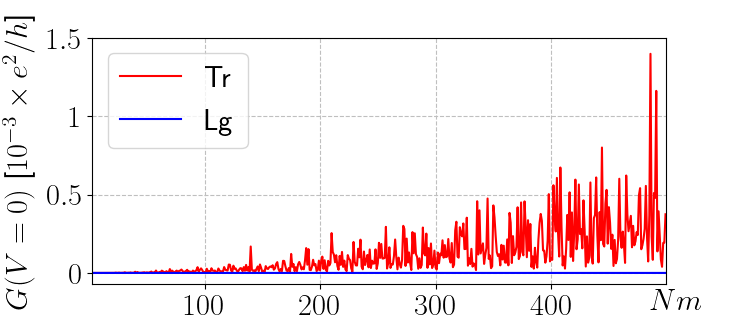}
        \subcaption{}
        \label{subfig:spin-valve-antiparallel-simulation-highscrew-vs-length}
    \end{subfigure}
    \begin{subfigure}[t]{0.49\textwidth}
        \centering
        \includegraphics[width=\textwidth]{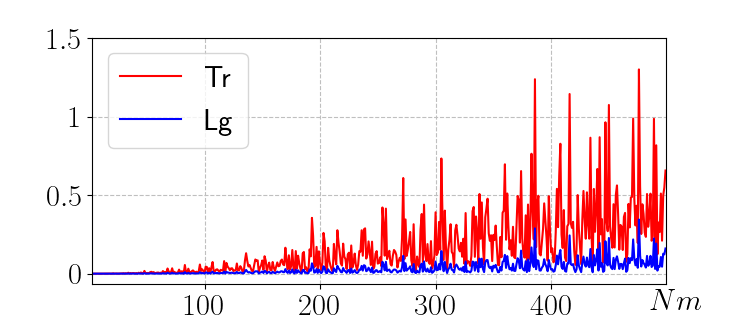}
        \subcaption{}
        \label{subfig:spin-valve-antiparallel-simulation-noscrew-vs-length}
    \end{subfigure}
    \\
    \begin{subfigure}[t]{0.49\textwidth}
        \centering
        \includegraphics[width=\textwidth]{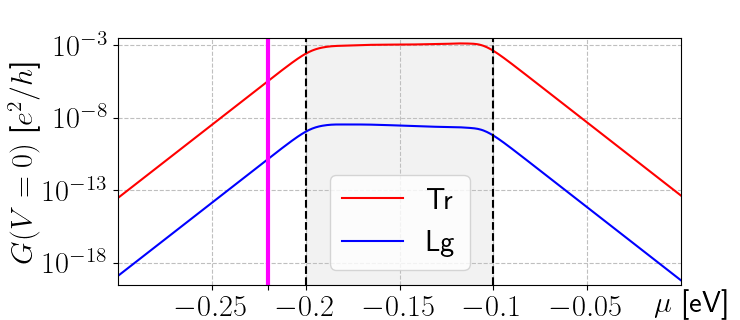}
        \subcaption{}
        \label{subfig:spin-valve-antiparallel-simulation-highscrew-vs-mu}
    \end{subfigure}
    \begin{subfigure}[t]{0.49\textwidth}
        \centering
        \includegraphics[width=\textwidth]{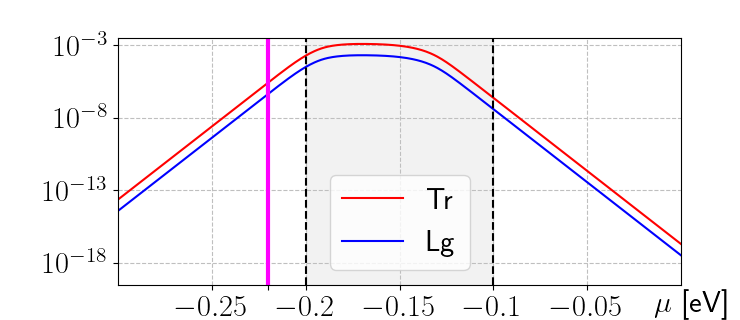}
        \subcaption{}
        \label{subfig:spin-valve-antiparallel-simulation-noscrew-vs-mu}
    \end{subfigure}
    \caption{\justifying Spin-valve transport simulations for a long molecule with screw symmetry (left) and weakly broken screw symmetry (right). (a) Sketch of an AP spin-valve setup. (b,c), (d,e), and (f,g) Current as a function of bias voltage, zero-bias conductance as a function of molecule length, and chemical potential, respectively, for a screw-symmetric molecule, $\gamma = 0$ (left) and weakly broken screw symmetry, $\gamma = 0.25$ (right). In (b), the Lg current does not exceed $10^{-4}\, \text{nA}$. In (f) and (g), the vertical axis is in logscale, the highlighted area indicates the conducting regime (where bands of the molecule and that of the leads overlap), and the solid purple vertical line indicates the chemical potential taken in (b,c,d,e). Parameters (if not varying) are $N=50$, $m=3$, $t_0=t_\mathrm{lead}=t_\mathrm{inter}=0.1$eV, $\varepsilon_Z=100t_0$, $\varepsilon_0=-\varepsilon_Z-3t_0$, $\mu=-2.2t_0$, $T=50$K, $\lambda=t_0/1000$, $\bxi_0 = (1, 1, 0.5)^T \times 2/3$. }
    \label{fig:spin-valve-simulation}
\end{figure*}

 In spintronics terms, this work predicts that the screw geometry of a molecule may greatly enhance the spin-mixing time, the typical time over which the spin of the electron is preserved when traversing the molecule.
A natural setup to test this prediction would be that of a spin-valve \cite{sanvito_molecular_2011} --- a two-terminal transport setup with two magnetic leads connected by the molecule, as is illustrated in Fig.~\ref{subfig:spin-valve-antiparallel-sketch}. 
An external field can be used to flip the magnetization orientation of the leads. We underline the difference with what is sometimes referred to as \emph{spin-valve} in CISS literature, where usually only one of the leads is magnetic \cite{bloom_chiral_2024}.

For simplicity, we consider that the leads' orbitals that inject and absorb electrons from the molecule are fully spin-polarized. 
If the leads are in \emph{parallel} (P) magnetization mode, then the spins of both leads are the same. Otherwise, in \emph{antiparallel} (AP) mode, the spins are opposite. In the former case, spin-flip processes inside the molecule would reduce the current, whereas in the latter case, spin-flipping processes are necessary to enable a nonzero current. We will focus on the AP setup, where the current can be used as a direct estimation of spin flips upon transport.

If the magnetization direction of the leads is aligned with the helical axis of the molecule (\emph{longitudinal}, or ``Lg'' mode), the spins are quantized along $\uvec_3$, spin flips are strongly suppressed, and, consequently, the current vanishes (in AP mode). Conversely, if the magnetization is along $\uvec_1$ (\emph{transverse}, or ``Tr'' mode), spin flips are allowed since spin quantized along $\uvec_1$ is not conserved in the molecule and a finite current will occur. 

\subsection{Transport simulations}
\label{sec:electron-transport-simulations}

We now turn to the numerical calculation of conductance of the spin-valve setup to support our expectations on experimental measurements.
The simulations are performed using the Kwant package \cite{groth_kwant_2014}. For the molecule, we use the model of Sec.~\ref{subsec:transition-from-translational} with the Hamiltonian given in Eq.~\eqref{eq:hamiltonian-imbalanced-helical-strand} with $\chi = 1$, $m=3$, and a finite number of coils $N \in \mathbb N^*$.
The sites on the right and left leads have a spin-dependent onsite energy $\varepsilon_X (\sigma)$ with the lead identifier $X \in \{\text{Right}, \text{Left}\}$ and a hopping amplitude $t_\text{lead}$, so that the dispersion relation in the leads reads $E_\text{lead} (k, \sigma, X) = - 2 t_\text{lead} \cos(k) + \varepsilon_X (\sigma)$. The energy $\varepsilon_X (\sigma)$ is decomposed as $\varepsilon_0 + s_X \sigma \varepsilon_Z$, with $\varepsilon_Z$ the intensity of the Zeeman field in the magnets, $s_\text{Right} = -s_\text{Left} = 1$ (using opposed fields in the two leads to implement the AP mode), and $\sigma=\pm 1$ the spin. The direction of the spin quantization axis in the leads is chosen parallel or perpendicular to that in the molecule for the Lg and the Tr mode, respectively. We typically take a very large $|\varepsilon_Z|$ to have fully spin-polarized orbitals and adjust $\varepsilon_0$ so that the energy of one of the orbitals in the leads matches that of the orbitals in the molecule. Lastly, the interfaces are represented by a hopping amplitude $t_\text{inter}$ between the leads and the molecule's peripheral sites.

For a set of parameters $t_\text{lead}, \varepsilon_0, \varepsilon_Z, t_0, \lambda, \bxi_0, \gamma, N, t_\text{inter}$, we compute the transmission probability $\tau (E)$ from the left to the right lead as a function of the mode's energy $E$. Note that $\tau (E)$ is sensibly nonzero only within the bandwidth of the molecule, eventually slightly overflowing as electrons may tunnel through it.

At a given voltage bias $V$, chemical potential $\mu$, and temperature $T$, we use the Landauer-Büttiker formalism to calculate the total current traversing the molecule \cite{datta_electronic_1995},
\begin{equation}
    I = -\frac e h \int \tau (E) \left(f \left(E + \frac{e V} 2\right) - f \left(E - \frac{e V} 2\right)\right) \mathrm d E,
\label{eq:landauer-buettiker}
\end{equation}
where $f(E)$ is the Fermi-Dirac distribution function,
\begin{equation}
    f (E) = \frac 1 {1 + e^{(E - \mu) / k_B T}}.
\end{equation}
Furthermore, by taking the $V \to 0$ limit of Eq.~\eqref{eq:landauer-buettiker}, we evaluate the conductance at zero voltage,
\begin{equation}
    G(V=0) = \left.\frac{\partial I}{\partial V}\right|_{V=0} = -\frac {e^2} h \int \tau(E) \frac{\partial f}{\partial E} \mathrm d E.
    \label{eq:zero-voltage-conductance}
\end{equation}

The results in Figs.~\ref{subfig:spin-valve-antiparallel-simulation-highscrew-vs-voltage},~\ref{subfig:spin-valve-antiparallel-simulation-noscrew-vs-voltage},~\ref{subfig:spin-valve-antiparallel-simulation-highscrew-vs-length},~\ref{subfig:spin-valve-antiparallel-simulation-noscrew-vs-length},~\ref{subfig:spin-valve-antiparallel-simulation-highscrew-vs-mu} and~\ref{subfig:spin-valve-antiparallel-simulation-noscrew-vs-mu} show the current-voltage characteristic and the length- and energy-dependence of the zero-bias conductance for a screw-symmetric molecule (left) and for weakly broken screw symmetry (right). A noticeable feature of these plots is that the Lg conformation hosts practically no current for a screw-symmetric molecule, whereas it does in Tr mode and/or when the screw symmetry is broken, as SOC can then induce spin flips. In Figs.~\ref{subfig:spin-valve-antiparallel-simulation-highscrew-vs-voltage} and~\ref{subfig:spin-valve-antiparallel-simulation-noscrew-vs-voltage}, the ratio between Tr and Lg currents is at least on the order of $10^3$ in the screw-symmetric case ($\gamma=0$) and drops below $10$ if the screw symmetry is slightly broken ($\gamma=0.25$). Also the zero-bias conductances shown in Figs.\ ~\ref{subfig:spin-valve-antiparallel-simulation-highscrew-vs-mu} and~\ref{subfig:spin-valve-antiparallel-simulation-noscrew-vs-mu} differ by approximately five orders of magnitude in both the insulating and conducting regimes for full screw symmetry. This is a striking evidence of the enhanced spin fidelity which could be measured experimentally.
We emphasize that this extinction of Lg current for $\gamma=0$ happens even if the direction of SOC, parametrized by $\bxi_0$, has components in the $(\uvec_1, \uvec_2)$ plane, which is precisely the consequence of screw symmetry. 

Lastly, we briefly discuss the dependence of $G(V=0)$ on the molecule's length $mN$, shown in Figs.~\ref{subfig:spin-valve-antiparallel-simulation-highscrew-vs-length} and~\ref{subfig:spin-valve-antiparallel-simulation-noscrew-vs-length}. The gradual increase in conductance can be attributed to the fact that in a short molecule, the dwell time is shorter than the spin-mixing time; the spin does not flip upon traversing the molecule and the particle entering from one spin-polarized lead reflects fully at the other, oppositely polarized lead. 
The length scale of this initial increase is thus $1/ \delta k$, where $\delta k$ is the SOC-induced quasi-momentum splitting. In a first order approximation, $\delta k$ is linear in $\lambda$, and thus this length scale is roughly proportional to $1/\lambda$. Fluctuations on a much shorter length scale visible in Figs.~\ref{subfig:spin-valve-antiparallel-simulation-highscrew-vs-length} and~\ref{subfig:spin-valve-antiparallel-simulation-noscrew-vs-length} can be attributed to the variation of the probability density. 
 
\subsection{Controlling the screw symmetry of a molecule}

The ideal experiment would measure transport properties of a screw-symmetric molecule under controlled symmetry-breaking perturbations, or at least those of two molecules with different screw conformations but comparable spin-splitting strengths. A possible platform could be provided by chiral soft matter, which enables molecular design on the nanometer scale. It is, for instance, possible to construct a helix of tuneable screw parameters using $\alpha$-aminoisobutyric acid (Aib) \cite{gatto_peptide_2022}, which has been recently successfully used as a CISS device \cite{rohmer_chiral_2025}. Twisted atomic wires of transition metal trichalcogenides \cite{shang_light-induced_2024} are another possible platform.

Interestingly, destroying screw symmetry by modifying onsite potentials while preserving translation invariance can be realized by placing an elongated object, such as a gate or another molecule, alongside one side of the helical molecule. In a film of upright, parallel molecules with a spatially varying intermolecular distance, the degree of screw symmetry would then be expected to depend on the gradient of this distance: screw symmetry is best preserved where the distance gradient vanishes and is progressively more strongly broken for larger gradients. 

\section{Discussion and conclusion}
\label{sec:conclusion}\label{sec:discussion}

This work reveals a mechanism of a robust, effective alignment of spin with the screw axis, in long, screw-symmetric molecules with small SOC. The spin alignment derives from the screw geometry with a screw angle far from a multiple of $2\pi$. 
The main consequence is the quasi-absence of spin-flip scattering in the bulk of the molecule. An experimental verification of this effect of enhanced spin fidelity is possible in spin-valve setups with two magnetic leads with antiparallel magnetization. Preserved spin inside the molecule for spins aligned with the screw axis lead to a difference of nearly 100\% in the conductance, for lead magnetizations along and transverse to the screw axis.

This result could have important implications for our understanding of CISS, as many chiral molecules are (nearly) screw-symmetric and feature very weak spin-dependent interactions. Indeed, the enhanced spin fidelity is consistent with mechanisms that amplify  {spin selectivity in transport, such as the close-to-the-barrier tunneling mechanism \cite{Baranov2026}, or through coupling to vibrations \cite{fransson_vibrational_2020,fransson_chiral_2025}. 
The present work shows that spin splitting stemming, e.g., from SOC or current-induced magnetic fields, is stabilized by screw symmetry along the helical axis, which is a necessary ingredient for most spin-selective transport mechanisms.}
 
In realistic setups, phonons and disorder could break the screw-symmetry-induced enhanced spin fidelity by allowing spin-flip scattering accompanied with a momentum change, $(k, \sigma) \to (k', -\sigma)$. 
However, in most materials, the energy of phonons is much smaller than the scale of the normal hopping amplitude. Consequently, the condition of small energy change (compared to band width), $E_{k, \sigma} \approx E_{k', -\sigma}$, is still intact, which is also trivially true for disorder scattering. As discussed in Sec.\ \ref{sec:spin-flip-derivation} and illustrated in Fig.\ \ref{fig:brillouin-split}, the distance between $k$ and $k'$ satisfying this condition is $\chi\theta$. One can thus envisage a spin-flip scattering $(k, \uparrow) \to (k + \chi \theta + \delta k, \downarrow)$ associated with a phonon momentum $\chi \theta + \delta k$, with $\delta k \ll 1$ sufficiently small to ensure that $E_{k, \uparrow} - E_{k + \chi \theta + \delta k, \downarrow}$ is of the order of the energy of a phonon. In the case of a discrete screw symmetry ($\theta \notin 2 \pi \mathbb Z$), the momentum of the phonon must be large. However, at low temperature, only small-momentum phonons are available. Hence, in screw-symmetric molecules, the enhanced spin fidelity is expected to be robust against phonon scattering at low temperatures. A similar robustness is expected with respect to disorder scattering if the disorder potential is smooth on the scale of the unit cell, in which case the momentum change upon scattering is also small as in the case of low-temperature phonons.

Another element that could spoil the spin fidelity is a spin-active interface, where the spin can flip upon transmission \cite{bloom_chiral_2024}. Controlling these so-called spinterface effects via chemically engineered interfaces \cite{ciudad_sign_2014} or considering the dependence of transport on the molecule length are possible ways to isolate the effect of enhanced spin fidelity due to screw symmetry.

On a conclusive note, it appears that two-terminal transport measurements with both leads being magnetic are less studied in the context of CISS. Without disregarding the experimental challenge it represents, our work demonstrates that such spin-valve setups can provide insightful perspectives on the interaction between spin and chirality, and the role of spin preservation due to screw symmetry.

\begin{acknowledgments}
We thank Robert Bittl and Virginia Gali for useful discussions. This work was supported by the Deutsche Forschungsgemeinschaft (DFG, German Research Foundation) - Project Number 277101999 - CRC-TR 183 (project A02), the Emmy Noether program - Project Number 506208038 -, and Cluster of Excellence EXC 3112 Center for Chiral Electronics.
\end{acknowledgments}

\bibliography{library}

\appendix

\section{Spin fidelity in a helical ion strand}
\label{apx:continuous-screw-symmetry}

Analyzing the model of Sec.~\ref{subsec:transition-to-continuum} leads us to the following base dispersion and eigenstates, expressed in terms of the spinful Wannier orbitals $\ket{j, \sigma} = \ket{j} \otimes \ket{\sigma}$:
\begin{align}
  E_{k, \sigma} &= - 2 t_0 \cos\left(k - \frac \pi m \chi \sigma\right), \\
  \ket{\psi_{k, \sigma}} &= \frac 1 {\sqrt{m N}} \sum_{j=0}^{m N - 1} e^{i j (k - \frac \pi m \chi \sigma)} \ket{j, \sigma}.
\end{align}
The Bloch Hamiltonian at momentum $k$ then takes the form
\begin{align}
  h(k) &= \begin{pmatrix}
  E_{k, \uparrow} + u_{k, \uparrow} & t_k^* \\
  t_k & E_{k, \downarrow} + u_{k, \downarrow}
\end{pmatrix}, \text{with} \\
    u_{k, \sigma} &= 2 \sigma \lambda \sin\left(2 \left(k - \frac\pi m \sigma \chi\right)\right) \bxi_0 \cdot \uvec_3 \text{ and}  \\
    t_k &= 2 e^{-i 2 \pi \chi / m} \lambda \sin(2 k) \bxi_0 \cdot (\uvec _1 + i \uvec _2). 
\end{align}
Now, we can compute the spin fidelity for the state $\ket{\psi_{k, \sigma}}$ as follows:
\begin{align}
    \ket{\psi(t = 0)} &= \ket{\psi_{k, \sigma}} = (c_{k, \sigma}^+)^* \ket{\psi_{k, +}} + (c_{k, \sigma}^-)^* \ket{\psi_{k, -}} \\
    \ket{\psi(t)} &= e^{-\frac i \hbar \varepsilon_k^+ t}(c_{k, \sigma}^+)^* \ket{\psi_{k, +}} + e^{-\frac i \hbar \varepsilon_k^- t} (c_{k, \sigma}^-)^* \ket{\psi_{k, -}} \\
    \begin{split}
    \mathcal F &= \min_t \left|\braket{\psi(t)}{\psi_{k, \sigma}}\right|^2 \\
    &= \left(|c_{k, \sigma}^+|^2 - |c_{k, \sigma}^-|^2\right)^2.
    \end{split}
\end{align}
In the expression above, we denoted $\ket{\psi_{k, \pm}}$ the new eigenstates of the Hamiltonian including $\hat H_{\text{sp}}$, and $\varepsilon_k^{\pm}$ their energies.
Then, the computation can be done using the identities $|c_{k, \sigma}^+|^2 + |c_{k, \sigma}^-|^2 = 1$ and Eq.~\eqref{eq:eigvecs-coefficients-ratio}. In the end, one finds, for both spins,
\begin{equation}
    \mathcal F = \frac{r_k} {1 + r_k} \text{ where } r_k = \left(\frac{E_{k, \up} - E_{k, \downarrow} + u_{k, \up} - u_{k, \downarrow}}{2|t_k|}\right)^2.
\end{equation}
This expression makes it clear that $r_k$ grows to $+\infty$ when $E_{k,\up} - E_{k,\downarrow}$ dominates the other energy scales, leading to a spin fidelity equal to $1$.

\section{Bloch Hamiltonian in the model of Sec.~\ref{subsec:transition-from-translational}}
\label{apx:translational}

We cannot rely on the screw symmetry anymore and must write a 3-dimensional Bloch Hamiltonian for deriving the base eigenstates. Then, the spinful Bloch Hamiltonian will be of size $6$, making an analytical derivation of its eigenstates impossible. We therefore resort to numerical evaluation of the spin fidelity.

The base eigenstates are obtained by diagonalizing the following Bloch Hamiltonian:
 \begin{equation}
 h_0 (k) = - t_0 \begin{pmatrix}
   0 & 1 + \gamma & e^{-i k} \\
   1 + \gamma & 0 & 1 - \gamma \\
   e^{i k} & 1 - \gamma & 0
 \end{pmatrix}.
 \end{equation}
 Doing this gives us coefficients $(c_{k}^{n, p})$ with $n, p \in \{0, 1, 2\}$ such that the eigenstates are
 \begin{equation}
 \ket{\psi_{k, n, \sigma}} = \frac 1 {\sqrt{N}} \sum_{j=0}^{N - 1} e^{i j k} \sum_{p=0}^2 c_k^{n, p} \ket{3j + p, \sigma},
 \end{equation}
 with energies denoted $E_k^n$. The full Bloch Hamiltonian, including $\hat H_\text{sp}$ is determined by the following matrix elements:
 \begin{align}
 \begin{split}
   &\bra{\psi_{k, n', \sigma}} \hat H \ket{\psi_{k, n, \sigma}} = \delta_n^{n'} E_k^n \\
   &+ i \sigma \lambda \bxi_0 \cdot \uvec_3  \sum_{p=0}^2 \left[\left(c_k^{n', p + 2}\right)^* c_k^{n, p} e^{-(1 - \delta_p^0)i k} \right. \\
   &\left.- \left(c_k^{n', p}\right)^* c_k^{n, p + 2} e^{(1 - \delta_p^0)i k}\right],
 \end{split} \\
 \begin{split}
   &\bra{\psi_{k, n', -\sigma}} \hat H \ket{\psi_{k, n, \sigma}} = i \lambda \bxi_0 \cdot (\uvec_1 + \sigma i \uvec_2)\sum_{p=0}^2 e^{i 2\pi p \chi \sigma / 3} \\
   &\times \left[\left(c_k^{n', p + 2}\right)^* c_k^{n, p} e^{-(1 - \delta_p^0)i k} - \left(c_k^{n', p}\right)^* c_k^{n, p + 2} e^{(1 - \delta_p^0)i k}\right].
 \end{split}
 \end{align}

Since the base system has now three bands instead of one, the spin fidelity ought to be computed numerically for each band. 

\end{document}